\RequirePackage{fix-cm}
\documentclass[twocolumn]{svjour3}          
\smartqed  
\usepackage{graphicx}
\usepackage{amsmath,amssymb,amsfonts}
 \journalname{J Supercond Nov Mag}
\begin{document}

\title{Mechanism for enhanced disordered screening in strongly correlated
metals: local vs. nonlocal effects}



\author{Eric C. Andrade \and Eduardo Miranda \and Vladimir Dobrosavljevi\'{c}
}


\institute{Eric C. Andrade \at
              Institut f\"{u}r Theoretische Physik, Technische Universit\"{a}t Dresden,
01062 Dresden, Germany 
           \and
           Eduardo Miranda \at
              Instituto de F\'{i}sica Gleb Wataghin, Unicamp, Campinas,
SP 13083-859, Brazil
\and
           Vladimir Dobrosavljevi\'{c} \at
              Department of Physics and National High Magnetic Field Laboratory,
Florida State University, Tallahassee, FL 32306, USA \\
              \email{vlad@magnet.fsu.edu}	
}

\date{Received: date / Accepted: date}

\maketitle

\begin{abstract}
We study the low temperature transport characteristics of a disordered
metal in the presence of electron-electron interactions. We compare
Hartree-Fock and dynamical mean field theory (DMFT) calculations to
investigate the scattering processes of quasiparticles off the screened
disorder potential and show that both the local and non-local (coming
from long-ranged Friedel oscillations) contributions to the renormalized
disorder potential are suppressed in strongly renormalized Fermi liquids.
Our results provide one more example of the power of DMFT to include
higher order terms left out by weak-coupling theories.
\end{abstract}

\section{Introduction}
\label{intro}

Transport in dirty metals has been investigated for many years, with
a substantial theoretical and experimental understanding being achieved
in the case of weak disorder and moderate electron-electron interaction
\cite{lee_ramakrishnan,punnose05}. However, much less is known
about situations with strong electronic correlations, where
much of our understanding relies on the application
of numerical methods like quantum Monte Carlo \cite{scalettar03}
or exact diagonalization \cite{scalettar08}, which are nevertheless
severely limited in temperature range and system sizes. A more flexible
approach to investigate the interplay of strong correlations and disorder
is provided by the dynamical mean field theory (DMFT) \cite{dmft_rmp96}.
In its original formulation, the DMFT treatment of disordered systems
does not include Anderson localization effects \cite{screening_2003},
a limitation which was quickly resolved by the introduction of the
statistical DMFT (statDMFT) \cite{statdmft,NFL_2005}. The statDMFT
approach has already led to some novel effects like a strong disorder screening
by interactions \cite{screening_2003,proceeding_sces08}, energy-resolved
spatial inhomogeneities \cite{proceeding_sces08}, and the emergence
of an Electronic Griffiths phase in the vicinity of the disordered
Mott transition \cite{egp01,griffiths2d09}. 

To partially elucidate the mechanism behind the rich physics uncovered through the
statDMFT method, a recent work \cite{ripples10} provided analytical
insights into the scattering off a weak random disorder
potential in an otherwise uniform strong\-ly inter\-act\-ing para\-mag\-net\-ic
metal. While the analysis is most straightforward and transparent
in this regime, this general issue is of key relevance also for the
diffusive regime. Here, we revisit this problem, explicitly comparing
the statDMFT results with those of the Hartree-Fock (HF) approximation. Our analytical
results  highlight the non-perturbative nature of the statDMFT and 
show that processes left out by HF generate vertex
corrections to the impurity potential, which ultimately lead to enhanced
screening \cite{screening_2003}. 

\section{Model and methods}
\label{sec:1}

We study the paramagnetic phase of the disordered Hubbard
model \begin{eqnarray}
\mathcal{H} & = & \sum_{i\sigma}\varepsilon_{i}n_{i\sigma}-\sum_{\left\langle ij\right\rangle ,\sigma}t_{ij}c_{i\sigma}^{\dagger}c_{j\sigma}+U\sum_{i}n_{i\uparrow}n_{i\downarrow},\label{eq:hub}\end{eqnarray}
where $t_{ij}$ are the hopping matrix elements between near\-est-neighbor
sites, $c_{i\sigma}^{\dagger}$$\left(c_{i\sigma}\right)$ is the
creation (annihilation) operator of an electron with spin projection
$\sigma$ at site $i$, $U$ is the on-site Hubbard repulsion, $n_{i\sigma}=c_{i\sigma}^{\dagger}c_{i\sigma}$
is the number operator, and $\varepsilon_{i}$ are the site energies
(bare disorder potential). We consider here its paramagnetic phase at half-filling
(chemical potential $\mu=U/2$) and a particle-hole symmetric lattice.
Below we discuss two different routes to treat this model. 

\subsection{Hartree-Fock (HF)}
\label{sec:1-1}

To solve the model \eqref{eq:hub}, we first
consider the HF approach, as described, for example, in Refs. \cite{herbut01,nandini04}.
Here, we simply decouple the interaction term in \eqref{eq:hub} as
$n_{i\uparrow}n_{i\downarrow}\approx\left\langle n_{i\uparrow}\right\rangle n_{i\downarrow}+n_{i\uparrow}\left\langle n_{i\downarrow}\right\rangle -\left\langle n_{i\uparrow}\right\rangle \left\langle n_{i\downarrow}\right\rangle $.
We restrict ourselves to the paramagnetic solution, $\left\langle n_{i\uparrow}\right\rangle =\left\langle n_{i\downarrow}\right\rangle =\left\langle n_{i}\right\rangle $,
and the self-consistency condition is obtained from
\begin{eqnarray}
\left\langle n_{i}\right\rangle =T\sum_{\omega_{n}}G_{ii}\left(\omega_{n}\right) & = & T\sum_{\omega_{n}}\left[\frac{1}{i\omega_{n}\mathbf{1}-\mathbf{v}-\mathcal{H}_{0}}\right]_{ii},\label{eq:hf_mf}\end{eqnarray}
where $T$ is the temperature, $G_{ii}\left(\omega_{n}\right)$ is
the local part of the lattice Green's function, $\omega_{n}$ are the Matsubara
frequencies, $\mathcal{H}_{0}$ is the clean ($\varepsilon_{i}=0$)
and non-interacting ($U=0$) lattice Hamiltonian, $\mathbf{v}$ is
a site-diagonal matrix $\left[\mathbf{v}\right]_{ij}=v_{i}\delta_{ij}$,
whose entries \begin{eqnarray}
v_{i}=\varepsilon_{i}+\Sigma_{i}\left(0\right)-\mu & = & \varepsilon_{i}+U\left\langle n_{i}\right\rangle -\mu\label{eq:v_i_hf_def}\end{eqnarray}
are the renormalized HF disorder potential, and $\Sigma_{i}\left(0\right)=\Sigma_{i}\left(\omega\right)=U\left\langle n_{i}\right\rangle $
is the frequency-independent HF electronic self-energy. We note
that the HF approximation can be regarded as the static (weak-coupling)
limit of the statDMFT and that it already contains one of its most important
features: a self-energy which is local, but which varies
from site to site reflecting spatial disorder fluctuations. As we will show later, the statDMFT
contains all the HF diagrams, but it also re-sums many higher order
terms left out by HF. 

In general, we have to solve the self-consistency equation in \eqref{eq:hf_mf}
numerically. However, for a weak disorder potential ($\left|\varepsilon_{i}\right|\ll D$, where $D$ is the bare half band width),
we can expand it around the uniform solution, and, to leading order
in the disorder potential, we have \begin{eqnarray}
v\left(\vec{q}\right) & = & \frac{\varepsilon\left(\vec{q}\right)}{1-U\Pi_{\vec{q}}}+\mathcal{O}\left[\varepsilon\left(\vec{q}\right)^{2}\right],\label{eq:v_ren_hf}\end{eqnarray}
where $\varepsilon\left(\vec{q}\right)$ is the inverse lattice Fourier
transform of the disorder potential $\varepsilon_{i}$. $\Pi_{\vec{q}}$
is the usual static Lindhard polarization function \cite{Mahan_many_body}. 

As $U\rightarrow0$, the renormalized disorder potential reads $v_{i}\simeq\varepsilon_{i}+U\Pi_{ij}\varepsilon_{j}$,
where $\Pi_{ij}$ is the lattice Fourier transform of $\Pi_{\vec{q}}$,
showing that the electrons scatter not only off the bare impurities,
but also off the long-ranged potential generated by the Friedel oscillations
(encoded in $\Pi_{ij}$). This result contains another general feature
of the statDMFT: the renormalized disorder potential acquires non-local
terms, which are absent in the original DMFT formulation \cite{screening_2003}.

\begin{figure}[t]
\includegraphics[scale=0.43]{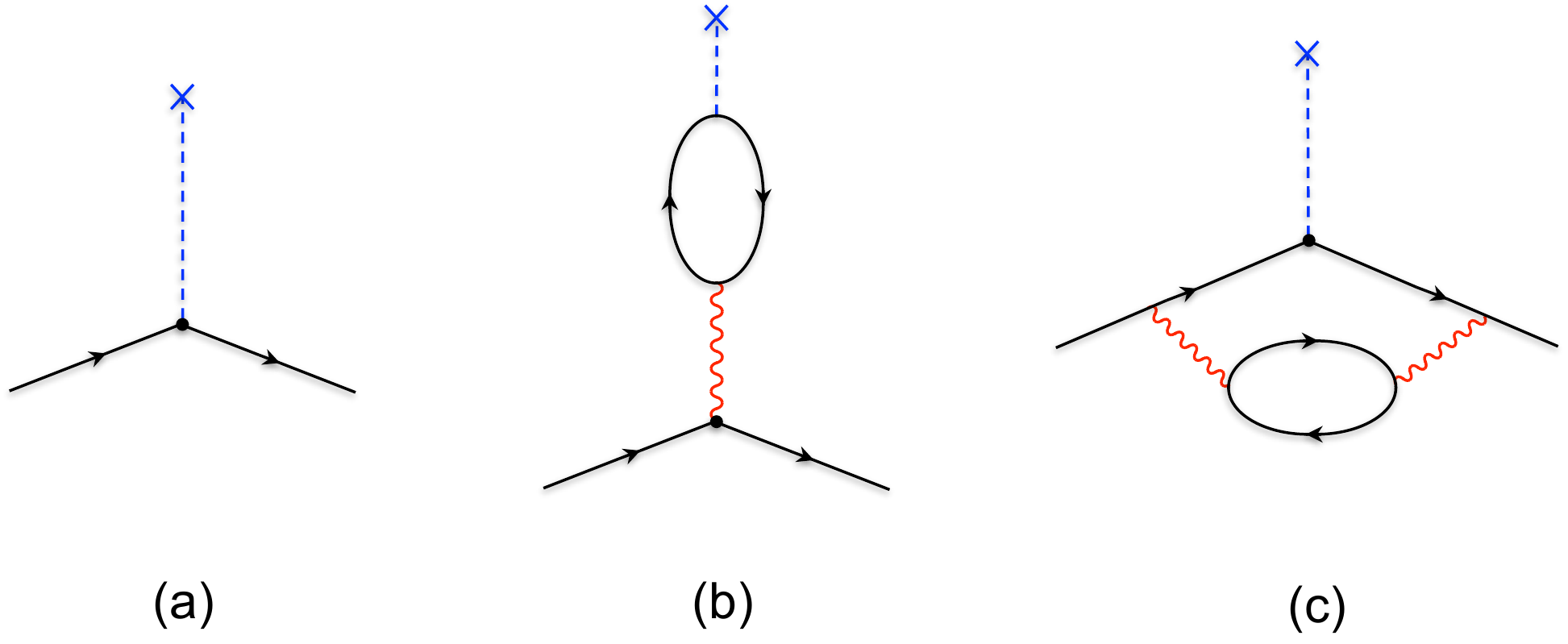}\caption{\label{fig:fig1} 
Diagrammatic representation of the
electron-impurity interaction: 
(a) the bare electron-impurity interaction vertex; 
(b) the first term of the Hartree-Fock theory, which is equivalent to 
the RPA resummation of ``bubbles''; 
(c) the first vertex correction to the electron-impurity interaction,
absent in Hartree-Fock theory, 
but included in the statDMFT/slave boson solution. 
The inclusion of these and higher-order 
vertex corrections is essential for the phenomenon
of perfect disorder screening. Here, the wavy line corresponds 
to the local on-site Hubbard-type electron-electron interaction.}

\end{figure}

We can easily interpret \eqref{eq:v_ren_hf} in terms of the usual
diagrammatic perturbation theory if we rewrite it as \begin{eqnarray}
\frac{v\left(\vec{q}\right)}{\varepsilon\left(\vec{q}\right)} & = & 
\frac{1}{\kappa_{RPA}(\vec{q})}=
1+U_{eff}^{HF}\left(\vec{q}\right)\Pi_{\vec{q}},\label{eq:v_ren_hf_1}\\
U_{eff}^{HF}\left(\vec{q}\right) & = & \frac{U}{1-U\Pi_{\vec{q}}} = \frac{U}{\kappa_{RPA}(\vec{q})},
\label{eq:Ueff_hf}\end{eqnarray}
where we have defined the dielectric function 
\begin{equation}
\kappa_{RPA}(\vec{q})=1-U\Pi_{\vec{q}},\label{eq:kappa_RPA}
\end{equation} 
which in this approximation is given by the RPA expression \cite{Mahan_many_body}.
It is then clear
that the HF approximation dresses the electron-impurity vertex by
a ``chain of bubbles'', as in the standard RPA screening theory \cite{Mahan_many_body},
Fig. (\ref{fig:fig1}b). 

\subsection{Slave-bosons (SB)}
\label{sec:1-2}

To investigate the strongly correlated regime,
we implement the statDMFT using the slave-boson (SB) mean-field theory
of Kotliar and Ruckenstein \cite{kotliar_ruckenstein} (which is equivalent
to the Gutzwiller variational approximation \cite{brinkman_rice})
as impurity solver \cite{proceeding_sces08,griffiths2d09}. This theory is mathematically
equivalent to applying directly the original formulation of Kotliar and Ruckenstein
to the Hubbard model in \eqref{eq:hub}, as discussed in Ref.  \cite{proceeding_sces08}.

As in the previous
HF calculation, we consider a weak disorder potential and expand the
relevant mean-field equations  \cite{proceeding_sces08,griffiths2d09}
around their uniform solution. For particle-hole symmetry $Z_{i}=Z_{0}+\emph{O}\left(\varepsilon_{i}^{2}\right)$
and we have, at $T=0$ \cite{ripples10} \begin{eqnarray}
v\left(\vec{q}\right) & = & \frac{\varepsilon\left(\vec{q}\right)}{1-u^2-\tilde{U}\Pi_{\vec{q}}}+\mathcal{O}\left[\varepsilon\left(\vec{q}\right)^{2}\right],\label{eq:v_ren_sb}\end{eqnarray}
where $u=U/U_{c}$ and 
\begin{equation}
\tilde{U}=\left(\frac{u}{2}-\frac{1}{1-u}\right)U.
\end{equation}
Eq. \eqref{eq:v_ren_sb} implies the following dielectric function
\begin{eqnarray}
\kappa_{SB}(\vec{q})&=&1-u^2-\tilde{U}\Pi_{\vec{q}}.\label{eq:kappa_sb}
\end{eqnarray}
The relation $n\left(\vec{q}\right)=1/2+\Pi_{\vec{q}}v\left(\vec{q}\right)$
also holds, as expected. It is instructive to consider the limits of weak and strong interactions.
If $U\ll U_c$,
\begin{equation}
\kappa_{SB}(\vec{q})\approx 1-U\Pi_{\vec{q}} = \kappa_{RPA}(\vec{q}),
\label{eq:kappa_sbweak}
\end{equation}
and we recover the HF result. In contrast, as $U \to U_c$,
\begin{equation}
\kappa_{SB}(\vec{q})\to -\frac{U_c}{1-u} \Pi_{\vec{q}},
\label{eq:kappa_sbstrong}
\end{equation}
leading to 
\begin{equation}
v_{i}\simeq-\left(1-U/U_{c}\right)$ $U_{c}^{-1}\left[\mbox{\boldmath\ensuremath{\Pi}}^{-1}\right]_{ij}\varepsilon_{j}.
\label{eq:v_ren_strong}
\end{equation}
As the system approaches the Mott transition, the renormalized disorder potential
goes to zero at all lattice sites, a situation that was dubbed ``perfect disorder screening'' in Ref. \cite{screening_2003}.
Its spatial structure is also very interesting, since $v_{i}$ is
just as non-local as for small $U$, but the non-local term is governed
not by the Lindhard function, but \emph{by its inverse}. 
The spatial structure of the charge disturbance in this limit is given by
\begin{equation}
\delta n_{i}=\left[\left(1-U/U_{c}\right)/U_{c}\right]\varepsilon_{i}+\mathcal{O}\left(\left(1-U/U_{c}\right)^{3}\right).
\label{eq:charge_sbstrong}
\end{equation}
Thus, although the charge fluctuations are suppressed everywhere,
its non-local part, coming from the Friedel oscillations,
is much more strongly suppressed ($\mathcal{O}\left(1-u\right)^{3}$)
and the electronic density is significantly disturbed only in the immediate
vicinity of the impurities. The suppression of the slow spatial decay
in $\delta n_{i}$ reflects the fundamental tendency of quasiparticles
to become localized as the system approaches the Mott insulator. Therefore,
density fluctuations are healed very effectively in the strongly correlated limit.

Additional insight into these results can be obtained by noting that the
second term on the right-hand side of Eq. \eqref{eq:kappa_sb} is unimportant both 
in the weakly and in the strongly correlated limits, \textit{cf.} Eqs. \eqref{eq:kappa_sbweak}
and \eqref{eq:kappa_sbstrong}. Neglecting this term in  \eqref{eq:kappa_sb}, we can
follow the same procedure as in \eqref{eq:v_ren_hf_1} and
rewrite \eqref{eq:v_ren_sb} as 
\begin{eqnarray}
\frac{v\left(\vec{q}\right)}{\varepsilon\left(\vec{q}\right)} & = & 1+U_{eff}^{SB}\left(\vec{q}\right)
\Pi_{\vec{q}},\label{eq:v_ren_sb_1}\\
U_{eff}^{SB}\left(\vec{q}\right) & = & \frac{\tilde{U}}{1-\tilde{U}\Pi_{\vec{q}}}.
\label{eq:Ueff_sb}
\end{eqnarray}
The approach to Mott localization in this language can thus be described by the replacement $U\to \tilde{U}$, 
\textit{cf.} Eqs. \eqref{eq:v_ren_hf_1} and \eqref{eq:Ueff_hf}.
This replacement, in turn, may be viewed as a \emph{local field correction} coming from vertex
corrections in the polarization function \cite{Mahan_many_body,loc_field05}
(see Fig. (\ref{fig:fig1}c) and the discussion in Section \ref{sec:3} below). 
Close to the Mott transition $\tilde{U}\approx U/(1-u)$ diverges and this strong correlation effect is seen 
to fall completely outside the scope of the HF theory. In fact, whereas HF predicts a gradual decrease 
of the dielectric function with increasing $U$, signaling the suppressed screening, see Eq. \eqref{eq:kappa_RPA},
the statDMFT/SB approach predicts precisely the opposite: the dielectric function diverges as $U \to U_c$, 
see Eq. \eqref{eq:kappa_sbstrong}, and the screening becomes \emph{asymptotically perfect}.
\section{Beyond weak-coupling} 
\label{sec:3}

Comparing the renormalized disorder potential
in \eqref{eq:v_ren_hf} and \eqref{eq:v_ren_sb}, we see that the
interaction corrections left out by HF generate vertex corrections
to the impurity potential, which are contained in the effective interaction
$U_{eff}^{SB}(\vec{q})$ in \eqref{eq:Ueff_sb}. Since the HF approximation
is the first term in expanding the electronic self-energy in $U$,
we expand our solutions \eqref{eq:v_ren_hf} and \eqref{eq:v_ren_sb}
in powers of $U$, in order to track down which terms are left out of
HF. As we have seen, to first order in $U$, the statDMFT/SB and HF solutions agree.
At second order, a difference  emerges already \begin{eqnarray}
\frac{v_{HF}^{\left(2\right)}\left(\vec{q}\right)}{\varepsilon\left(\vec{q}\right)} & = & U^{2}\Pi_{\vec{q}}^{2},\label{eq:vi_u0_hf}\\
\frac{v_{SB}^{\left(2\right)}\left(\vec{q}\right)}{\varepsilon\left(\vec{q}\right)} & = & U^{2}\Pi_{\vec{q}}^{2}+\left(\frac{U}{U_{c}}\right)^{2}\left(1+\frac{3}{2}U_{c}\Pi_{\vec{q}}\right).\label{eq:vi_u0_sb}\end{eqnarray}

To gain insight into the leading correction beyond HF, we combine the
statDMFT procedure with usual perturbation theory. First, we recall
that in the statDMFT approach the electronic self-energy is local, albeit site-dependent. The only contribution
to a local self-energy which is of second order in the interactions is given
by \cite{mueller-hartmann} 
\begin{eqnarray}
\Sigma_{i}^{\left(2\right)}\left(i\omega_n\right) & = & U^{2}T\sum_{\nu_n}
G_{ii}^{\left(0\right)}\left(i\omega_n+i\nu_{n}\right)\Pi_{ii}^{\left(0\right)}\left(i\nu_n\right),\label{eq:self_2_def}\end{eqnarray}
where \begin{eqnarray}
\Pi_{ii}\left(i\nu_n\right) & = & T\sum_{\nu^\prime_n}G_{ii}^{\left(0\right)}\left(i\nu^\prime_n\right)G_{ii}^{\left(0\right)}\left(i\nu^\prime_n-i\nu_n\right)\label{eq:pol_loc}\end{eqnarray}
is the local contribution of the dynamical Lindhard polarization, calculated
using the local Green's function $G_{ii}^{\left(0\right)}\left(i\omega_n\right)=\left[\left(i\omega_n-\mbox{\boldmath\ensuremath{\varepsilon}}-\mathcal{H}_{0}\right)^{-1}\right]_{ii}$
with $\mbox{\boldmath\ensuremath{\varepsilon}}_{ij}=\varepsilon_{i}\delta_{ij}$.
We are interested only in the leading $\omega_n$ behavior of \eqref{eq:self_2_def},
since our statDMFT/SB approach is itself a low-energy one \cite{vollhardt_he3}.
Ultimately, this low-energy approximation provides a local Fermi-liquid
description of the auxiliary impurity problem \cite{Hewson_kondo}. For example, the
uniform contribution $\left(\varepsilon_{i}=0\right)$ of \eqref{eq:self_2_def}
is  $\Sigma_{0}^{\left(2\right)}\left(i\omega_n\right)\approx$\linebreak  $\left(1-1/Z_{0}^{\left(2\right)}\right)i\omega_n + \mathcal{O}\left(\omega_n^{2}\right)$,
with $\left(1-1/Z_{0}^{\left(2\right)}\right)\propto-U^{2}$, for a particle-hole symmetric lattice. 

We now expand \eqref{eq:self_2_def} and \eqref{eq:pol_loc} to linear order in the impurity
potential. There are three identical contributions, each with one of the three Green's function
lines with an impurity vertex inserted in it, as shown in Fig (\ref{fig:fig1}c).
We focus on $\Sigma_{i}^{\left(2\right)}\left(0\right)$
since this defines the renormalized disorder potential. Qualitatively,
it is very easy to see how the extra terms in \eqref{eq:vi_u0_sb}
are generated. Consider, for simplicity, that we estimate $\Pi_{ii}\left(i\nu_n\right)$ in \eqref{eq:self_2_def}
through the clean and static limit $\Pi_{ii}\left(i\nu_n\right)\approx\Pi_{0}\left(0\right)=-\rho\left(0\right)
\approx -U_{c}^{-1}$.
In this case, we simply have $\Sigma_{i}^{\left(2\right)}\left(0\right)\approx\left(U/U_{c}\right)^{2}U_{c}
\Pi_{ij}\varepsilon_{j},$
which has the same structure as the last term in \eqref{eq:vi_u0_sb}.
Based on
these arguments, we stress that the difference which exists already
at order $U^{2}$ between \eqref{eq:vi_u0_hf} and \eqref{eq:vi_u0_sb}
is an interaction-generated vertex correction
of the electron-impurity vertex, which is absent in HF/RPA screening theory,
but which is re-summed to all orders within statDMFT/SB.

\section{Conclusions}
\label{sec:conclusions}

We presented a detailed analytical calculation
of the effects of weak disorder scattering in a correlated host. Comparing
the results obtained within HF and statDMFT, we highlighted the
fact that statDMFT incorporates important vertex corrections to all
orders, a task which is difficult, or more likely, even impossible
to perform using weak-coupling diagrammatic approaches. 
A physical consequence of the inclusion of these vertex corrections
is the phenomenon of disorder screening by interactions.

An analogous example of this dichotomy can be observed in the familiar
Migdal-Eliashberg (ME) strong coupling
theory describing the electron-phonon problem \cite{Mahan_many_body}. Indeed, the ME theory
neglects the momentum dependence of the electronic self-energy and
may thus be regarded as a weak-coupling approximation to DMFT. 
Like the HF/RPA screening described above, it also neglects vertex corrections.
The full DMFT solution, however, 
not only contains all the ME diagrams, but it
also re-sums many higher order terms left out by the ME approach, 
including vertex corrections \cite{hewson02prl}, 
in close analogy with the statDMFT treatment of disorder and interactions. In both models, these strong coupling effects reflect non-perturbative Kondo-like processes \cite{screening_2003,hewson02prl}, which cannot be described by weak-coupling approaches. 

\begin{acknowledgements}
This research was supported by the DFG through
FOR 960 and GRK 1621 (ECA),
by FAPESP through grant 07/57630-5 (EM), 
CNPq through grant\linebreak 304311/2010-3 (EM), and by the NSF through grant DMR-0542026 (VD). The authors thank Lev Gor'kov for pointing out the role of vertex corrections in the Holstein model.
\end{acknowledgements}

\appendix

\bibliographystyle{spphys}
\bibliography{proc}

\end{document}